\documentclass[a4paper,10pt]{article}
\usepackage{multirow}
\usepackage{graphicx}
\usepackage{axodraw4j}
\usepackage{amssymb}
\usepackage{amsmath}
\usepackage{cite}
\usepackage{graphicx,subfigure,epsfig,epsf}
\def\be{\begin{equation}}
\def\ee{\end{equation}}
\def\bea{\begin{eqnarray}}
\def\eea{\end{eqnarray}}

\def\issue(#1,#2,#3){#1 (#3) #2} 
\def\APP(#1,#2,#3){Acta Phys.\ Polon.\ \issue(#1,#2,#3)}
\def\ARNPS(#1,#2,#3){Ann.\ Rev.\ Nucl.\ Part.\ Sci.\ \issue(#1,#2,#3)}
\def\CPC(#1,#2,#3){Comp.\ Phys.\ Comm.\ \issue(#1,#2,#3)}
\def\CIP(#1,#2,#3){Comput.\ Phys.\ \issue(#1,#2,#3)}
\def\EPJC(#1,#2,#3){Eur.\ Phys.\ J.\ C\ \issue(#1,#2,#3)}
\def\EPJD(#1,#2,#3){Eur.\ Phys.\ J. Direct\ C\ \issue(#1,#2,#3)}
\def\IEEETNS(#1,#2,#3){IEEE Trans.\ Nucl.\ Sci.\ \issue(#1,#2,#3)}
\def\IJMP(#1,#2,#3){Int.\ J.\ Mod.\ Phys. \issue(#1,#2,#3)}
\def\JHEP(#1,#2,#3){J.\ High Energy Physics \issue(#1,#2,#3)}
\def\JPG(#1,#2,#3){J.\ Phys.\ G \issue(#1,#2,#3)}
\def\MPL(#1,#2,#3){Mod.\ Phys.\ Lett.\ \issue(#1,#2,#3)}
\def\NP(#1,#2,#3){Nucl.\ Phys.\ \issue(#1,#2,#3)}
\def\NIM(#1,#2,#3){Nucl.\ Instrum.\ Meth.\ \issue(#1,#2,#3)}
\def\PL(#1,#2,#3){Phys.\ Lett.\ \issue(#1,#2,#3)}
\def\PRD(#1,#2,#3){Phys.\ Rev.\ D \issue(#1,#2,#3)}
\def\PRL(#1,#2,#3){Phys.\ Rev.\ Lett.\ \issue(#1,#2,#3)}
\def\PTP(#1,#2,#3){Progs.\ Theo.\ Phys. \ \issue(#1,#2,#3)}
\def\RMP(#1,#2,#3){Rev.\ Mod.\ Phys.\ \issue(#1,#2,#3)}
\def\SJNP(#1,#2,#3){Sov.\ J. Nucl.\ Phys.\ \issue(#1,#2,#3)}

\def\d{\delta}
\def\ep{\epsilon}
\def\no{\nonumber}
\def\uno{\mbox{1 \kern-.59em {\rm l}}}
 
\begin{document}
\bibliographystyle{unsrt}
\begin{titlepage}

\vskip2.5cm
\begin{center}
\vspace*{5mm}
{\huge \LARGE Using an intense laser beam in interaction with muon/electron beam to probe the Noncommutative QED }
\end{center}
\vskip0.2cm

\begin{center}
{\it S. Tizchang$^\dag$, S. Batebi$^\dag$, M. Haghighat$^\ddag$ and R. Mohammadi$^\S$\footnote{rmohammadi@ipm.ir}}

\end{center}
\vskip 8pt

\begin{center}
$^\dag$ {\it \small Department of Physics, Isfahan University of Technology(IUT),\\ Isfahan, Iran 84156-83111.}\\
$^\ddag${\it  Department of Physics, Shiraz University ,\\Shiraz, Iran}\\
 $^\S${\it \small Iran Science and Technology Museum (IRSTM), PO BOX: 11369-14611, Tehran, Iran.}\\
\vspace*{0.3cm}


\end{center}


\begin{abstract}
It is known that the linearly polarized photons can partly transform to circularly polarized ones via forward Compton scattering in a background such as the external magnetic field or noncommutative space time.  Based on this fact we explore the effects of the NC-background on the scattering of a linearly polarized laser beam from an intense beam of charged leptons.  We show that for a muon/electron beam flux $\bar\varepsilon_{\mu,e}\sim 10^{12}/10^{10}\,{\rm TeV}\,{\rm cm}^{-2}\,{\rm sec}^{-1}$ and a linearly polarized laser beam with energy $k^0\sim $1 eV and average power $\bar{P}_{\rm laser}\simeq$1 MW, the generation rate of circularly polarized photons is about $R_{_V} \sim 10^4/{\rm sec}$ for Noncommutative energy scale $\Lambda_{\tiny{NC}}\sim 10$TeV.
 This is fairly large and can grow for more intense beams in near future.
\end{abstract}

\end{titlepage}
\section{INTRODUCTION}
The Compton scattering is one of the most important scattering in physics particularly in the cosmology, astrophysics and astro-particle physics. Although this scattering can be the  main source to generate a linear polarized wave, it cannot generate circular polarized wave. In fact, there is no many ways to generate circular polarization in a scattering process in the QED. Therefore, any experiment including the measurement of circular polarization can provide a way to understand more accurately the physics of scattering. It is shown that in the presence of a background field such as Noncommutativity (NC) in space-time or external magnetic field, the generation of circular polarization by the Compton scattering is possible\cite{bavarsad}.  Meanwhile, it is recently shown that the photons of a linearly polarized laser beam can acquire circular or B-mode linear polarization by scattering off active neutrinos via a parity violation process. This phenomenon can possibly be used to understand the nature of neutrinos (Dirac or Majorana), to detect the fluxes of Cosmic Neutrino Background \cite{Mohammadi} and to gain some insight into the physics of active and light sterile neutrino oscillations \cite{pouya}.   Here we would like to consider NC-space as a nontrivial background to explore the polarization of photons in the Compton scattering.  The NC field theory and its phenomenological aspects have been considered for many years without any sign in nature\cite{NC}. The NC space-time  which is predicted in the string theory should have a scale $\Lambda_{NC}$ of the order of the Planck scale.  Nevertheless, it is believed that $\Lambda_{NC}$ like the other new physics scales may has some impact at the TeV scale.   However, the null results mean that the parameter of noncommutativity, if exist, is minuscule.  In fact, to find the NC effects one should consider those processes in which such effects can be enhanced.
To this end we consider those experiments which are based on the laser technologies.  As in the Compton scattering in the NC-space the photons with the linear polarization transform partly to circularly polarized photons with a rate proportional to the NC-parameter, number of photons per pulse and the flux of fermions, one expects the enhancement of the rate of generation of the circular polarization with the intensities.
 Usually intense laser beams are designed  for a wide-ranging experimental program in fundamental physics and advanced applications such as the interaction of super high intensity light with matter, fast ignition fusion research, photon induced nuclear reactions, electron and ion acceleration by light waves, astrophysics in the laboratory and the exploration of the exotic world of plasma dominated by relativity. Some of the important laser beams are:
X-ray free-electron laser (XFEL) facilities \cite{XFEL}, optical high-intensity laser facilities such as Vulcan \cite{vulcan}, peta-watt laser beam \cite{peta} and ELI \cite{eli}, as well as SLAC E144 using nonlinear Compton scattering \cite{burke1997}.\par

In this work, we consider the forward Compton scattering through the collision of photons in a laser beam  with an intense and high energy  muon/electron beam in the noncommutative space to examine the generation of circular polarization.  Consequently, we calculate the generation rate of the circularly polarized photons $R_{_V}$ to explore the possibility of discovering the NC effects in different energy scales. It is clear to obtain a large value for $R_{_V}$ to be measurable in lab, one needs intense muon/electron and laser beams.  Therefore, an accelerator that can produce ultra-intense beams of muons and even electrons provides opportunities to discover the NC effect.\par
In Sec.II we review the Stokes parameters and Boltzmann equation formalism. In Sec. III the Noncommutative standard model (NCSM) is  briefly explained and the time evolution of the Stokes parameters related to the Compton scattering in NCSM is discussed. In Sec. IV  we study the effect of space-time Noncommutativity on the collision of a laser beam with an ultra-intense beam of muon/electron. In Sec.V we give an estimation on the generation rate of the circular polarization in such interactions. In Sec.VI some concluding remarks are given.

\section{Stokes parameters and Boltzmann equation }
Laser beam polarization can be characterized by four known Stokes parameters which are $I$ the total intensity of the beam, $Q$  and $U$ the linear polarization of the radiation and $V$ the difference between positive and negative circular polarization.
Moreover, the density operator of an ensemble of photons in terms of the Stokes parameters is defined as follows
\begin{eqnarray}
\hat\rho=\frac{1}{\rm {tr}(\hat \rho)}\int\frac{d^3\textbf{k}}{(2\pi)^3}
\rho_{ij}(\textbf{k})D_{ij}(\textbf{k}),\quad \rho=\frac{1}{2}\left(\begin{array}{cc}
             I+Q& U-iV \\
             U+iV & I-Q \\
                 \end{array}
        \right),\label{t0}
\end{eqnarray}
where
\begin{eqnarray}
I&=&\rho_{11}+\rho_{22},\label{i}\\
Q&=&\rho_{11}-\rho_{22},\label{q}\\
U&=&\rho_{12}+\rho_{21},\label{u}\\
V&=&i(\rho_{12}-\rho_{21}).
\label{v}
\end{eqnarray}
$\rho_{ij}(\textbf{k})$ is the general density-matrix which is related to the photon number operator $D^0_{ij}(\textbf{k})\equiv a_i^\dag (\textbf{k})a_j(\textbf{k})$ where $a_i^\dagger(\textbf{k})$ and $a_j (\textbf{k})$ are photon creation
and annihilation operators.
 The expectation value of the number operator is defined as
\begin{eqnarray}
\langle\, D^0_{ij}(\textbf{k})\,\rangle\equiv {\rm tr}[\hat\rho
D^0_{ij}(\textbf{k})]=(2\pi)^3 \delta^3(0)(2k^0)\rho_{ij}(\textbf{k}).\label{t1}
\end{eqnarray}
The time evolution of the operator $D^0_{ij}(\textbf{k})$, considered in the Heisenberg picture, is
\begin{equation}\label{heisen}
   \frac{d}{dt} D^0_{ij}(\textbf{k})= i[H,D^0_{ij}(\textbf{k})],
\end{equation}
here $H$ is the full Hamiltonian. Meanwhile, the time evolution equation for the density matrix up to the first order of the interacting Hamiltonian $H^0_I(t)$ for photons can be written as
\begin{eqnarray}
(2\pi)^3 \delta^3(0)(2k^0)
\frac{d}{dt}\rho_{ij}(\textbf{k}) &=& i\langle \left[H^0_I
(t);D^0_{ij}(\textbf{k})\right]\rangle \nonumber \\
&-&\frac{1}{2}\int dt\langle
\left[H^0_I(t);\left[H^0_I
(0);D^0_{ij}(\textbf{k})\right]\right]\label{bo}\rangle,
\hspace{1cm}
\label{density}
\end{eqnarray}
where the first term on the right-hand side is a forward scattering term, and the second one is a higher order collision term, usual scattering cross section. The leading-order interacting Hamiltonian for photon-charged fermion is generally given by \cite{Kosowsky}
\begin{eqnarray}
  H^0_I &=& \int d\mathbf{q} d\mathbf{q'} d\mathbf{p} d\mathbf{p'} (2\pi)^3\delta^3(\mathbf{q'} +\mathbf{p'} -\mathbf{p} -\mathbf{q} ) \nonumber \\
   &\times& \exp[it(q'^0+p'^0-q^0-p^0)]\left[b^\dagger_{r'}a^{\dagger}_{s'}(\mathcal{M})a_sb_r\right],\label{h0}
   \label{H}
\end{eqnarray}
where $\mathcal{M}$ is the amplitude of scattering matrix, $b_{r'}^\dagger(\textbf{k})$ and $b_r (\textbf{k})$ are charged fermion creation
and annihilation operators and $d\mathbf{q}\equiv \frac{d^3{\bf q}}{(2\pi)^3}\frac{m_f}{q^0}$ and $d\mathbf{p}\equiv \frac{d^3{\bf p}}{(2\pi)^32p^0}$, with the same relations for  $d\mathbf{q'}$ and $d\mathbf{p'}$,
 respectively.\par
 The operator expectation values are given as a function of number density of fermions per unit volume, $n_f(\textbf{q})$, where $\textbf{q}$ denotes the momentum of fermions\cite{Kosowsky}.
 Then the energy density and pressure of fermions are defined as
\begin{equation}\label{nd-n0}
\epsilon_f(\mathbf{x})=g_{f}\int \frac{d^3 \bf{q}}{(2\pi)^3}~q^0 \,\,n_f(\mathbf{x},\mathbf{q})~~~,~~~~ P_{f}=g_{f} \int\frac{ d^3\bf{q} }{(2\pi)^3}\, \frac{|\mathbf{q}|^2}{3~q^0}\,n_f(\mathbf{x},\mathbf{q}),
\end{equation}
 with $g_f$ shows the number of spin states or degrees of freedom.\par

\section{Time evolution of Stokes parameters vs NC Forward photon-fermion scattering}
In this section we explore the time evolution of the Stokes parameters via the forward photon-fermion scattering in the NC standard model.  The noncommutative version of standard model is introduced by two different approaches. In the first approach which is based on the "Seiberg-Witten" maps, the symmetry group, number of gauge fields and particles are the same as the ordinary standard model \cite{Connes}. Meanwhile, in the second approach the gauge group is $U(3)\times U(2) \times U(1)$ which can be reduced to the standard model gauge group through appropriate symmetry breaking \cite{Sheikh-Jabbari,Chaichian}. Here we follow the  first approach to do our calculations.\par
In the both versions of the noncommutative standard model, the ordinary coordinates convert to the noncommutative operators with a canonical form as
\begin{equation}
\left[\hat{x}^\mu,\hat{x}^\nu\right]=i\theta^{\mu\nu} \varpropto i\frac{1}{\Lambda_{\tiny{NC}}^2},
\end{equation}
where $\theta^{\mu\nu}$ is the parameter of noncommutativity and $\Lambda_{\tiny{NC}}=\frac{1}{\sqrt{|\theta^{\mu\nu}|}}$ is the noncommutative scale of energy. Also in the both version, Moyal-star product realization of the algebra are considered.
The noncommutative parameter, $\theta^{\mu\nu}$, can be usually divided into two parts: the time-space component $\theta^{0i}$ and
the space-space component $\theta^{ij}$ where $i,j,k=1,2,3$.  The NC scale for the space-space and the time-space parts can be defined as \cite{Haghighat}:
   \begin{equation}\label{theta}
\theta^{0i}=\frac{1}{\Lambda_{\tiny{TS}}^2} \hat{w}^{i},\,\,\,\,\,\theta^{ij}=\frac{1}{2~\Lambda_{\tiny{SS}}^2} \epsilon^{ijk}\hat{v}_{k}.
\end{equation}
  \par
where $w^i$ and $v^i$ refer to fixed direction in space. Although the time-space component of the NC parameter has some problems with the unitarity, the quantum mechanics can become unitary in some cases for the time-like part of NC-parameter too \cite{Balachandran,Pinzul}.\par
In the both approaches besides the corrections on the usual vertices new couplings also appear in comparison with the ordinary standard model.\par
\begin{figure}
\begin{picture}(55,45)(30,-30)
\SetWidth{0.5}
\hspace*{2cm}
\Photon(50,-25)(75,-50){2}{7}
\ArrowLine(50,-75)(75,-50)
\Vertex(75,-50){2.0}
\ArrowLine(75,-50)(105,-50)
\Photon(105,-50)(130,-25){2}{7}
\ArrowLine(105,-50)(130,-75)
\Text(45,-70)[lb]{$q$}
\Text(45,-40)[lb]{$p$}
\Text(80,-70)[lb]{$q+p$}
\Text(127,-40)[lb]{$p'$}
\Text(127,-70)[lb]{$q'$}
\end{picture}
\hspace*{5cm}
\begin{picture}(55,45) (30,-30)
\SetWidth{0.5}
\Photon(50,-25)(75,-50){2}{7}
\ArrowLine(50,-75)(75,-50)
\Vertex(105,-50){2.0}
\ArrowLine(75,-50)(105,-50)
\Photon(105,-50)(130,-25){2}{7}
\ArrowLine(105,-50)(130,-75)
\Text(45,-70)[lb]{$q$}
\Text(45,-40)[lb]{$p$}
\Text(80,-70)[lb]{$q+p$}
\Text(127,-40)[lb]{$p'$}
\Text(127,-70)[lb]{$q'$}
\end{picture}\\ \\
\hspace*{2cm}
\begin{picture}(55,45) (30,-30)
\SetWidth{0.5}
\Photon(110,-30)(75,-50){2}{7}
\ArrowLine(50,-75)(75,-50)
\Vertex(75,-50){2.0}
\ArrowLine(75,-50)(110,-50)
\Photon(110,-50)(75,-30){2}{7}
\ArrowLine(110,-50)(135,-75)
\Text(45,-70)[lb]{$q$}
\Text(65,-40)[lb]{$p$}
\Text(80,-70)[lb]{$q-p'$}
\Text(115,-40)[lb]{$p'$}
\Text(130,-70)[lb]{$q'$}
\end{picture}
\hspace*{3cm}
\begin{picture}(55,45) (30,-30)
\SetWidth{0.5}
\Photon(110,-30)(75,-50){2}{7}
\ArrowLine(50,-75)(75,-50)
\Vertex(110,-50){2.0}
\ArrowLine(75,-50)(110,-50)
\Photon(110,-50)(75,-30){2}{7}
\ArrowLine(110,-50)(135,-75)
\Text(45,-70)[lb]{$q$}
\Text(65,-40)[lb]{$p$}
\Text(80,-70)[lb]{$q-p'$}
\Text(115,-40)[lb]{$p'$}
\Text(130,-70)[lb]{$q'$}
\end{picture}\\ \\ \\ \\ \\ \\ \hspace*{6cm}
\begin{picture}(55,45) (30,-30)
\SetWidth{0.5}
\Photon(30,-30)(50,-10){2}{4}
\Photon(70,-30)(50,-10){2}{4}
\ArrowLine(30,11)(50,-10)
\ArrowLine(50,-10)(70,11)
\Vertex(50,-10){1.5}
\Text(40,11)[lb]{$q$}
\Text(77,8)[lb]{$q'$}
\Text(40,-40)[lb]{$p$}
\Text(77,-40)[lb]{$p'$}
\end{picture}
\vspace*{1.75cm}
\caption{Compton scattering in NC standard model}
\label{Compton}
\end{figure}
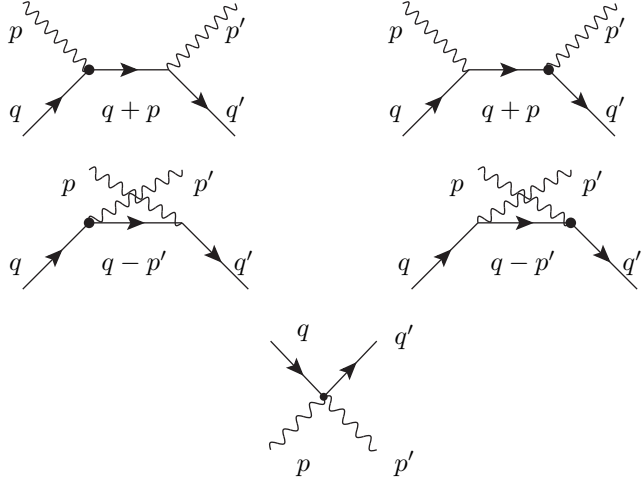

Using the NCSM based on the first approach \cite{Melic},  photon and charged fermion scattering can be described, up to the lowest order, by Feynman diagrams given in Fig(\ref{Compton}). Now we are ready to examine the time evolution of the Stokes parameters for photon-fermion scattering in NCQED.  It can be shown that the last diagram has not any contribution to the time derivative of circular polarization\cite{tizchang}. For the remaining diagrams given in Fig(\ref{Compton}) and using (\ref{H})
 one can find:
 \begin{eqnarray}
i\left<[H^0_I(0),\mathcal{D}_{ij}({\bf k})]\right>\!\!\!\!\!\!\!&&=-\frac{e^2Q_f^2}{2}\:(2\pi)^3\:\delta^{3}(0)\int
d{\bf q}\: n_f({\bf x},{\bf q})(\d_{is}\rho_{s'j}({\bf k})-\d_{js'}\rho_{is}({\bf k}))\no\\&&
\!\!\!\!\!\!\!\!\!\!\!\!\!\!\!\!\!\!\!\!\!\!\!\!\!\!\!\!\!\!\!\!\!\!\!\!\!\!\!\!\!\!\!\!\!\!\!
 \times
 \bar{u}_{r}(q)\{ \frac{1}{4 k\cdot q} \Big[q.\theta.\ep_{s'}(k)\Big(k\!\!\!\!/\:\:(q\!\!\!/+k\!\!\!/+m_f)\ep_{s}\!\!\!\!\!/\:\:(k)
 +\ep_{s}\!\!\!\!\!/\:\:(k)(q\!\!\!/-k\!\!\!/+m_f)
 k\!\!\!/\Big)\no\\&&
\hspace{-3cm}+q.\theta.\ep_{s}(k)\Big(\ep_{s'}\!\!\!\!\!/\:\:(k)(q\!\!\!/+k\!\!\!/+m_f)
 k\!\!\!/+k\!\!\!\!/\:\:(q\!\!\!/-k\!\!\!/+m_f)\ep_{s'}\!\!\!\!\!/\:\:(k)\Big)
\Big] \no\\&& \hspace{-3cm} + \Big[\ep_{s'}(k)\cdot
\theta\cdot \ep_s(k)\:\: k\!\!\!\!/\:\:- k \cdot\theta \cdot
\epsilon_s(k) \:\:\ep_{s'}\!\!\!\!\!/\:\:(k)\!\!\!\!\!\:\:\: + k
\cdot\theta \cdot \ep_{s'}(k)
\:\:\ep_{s}\!\!\!\!\!/\:\:(k)\Big] \}u_{r}(q). \hspace{1cm}
\end{eqnarray}
and consequently one has
\begin{eqnarray}
\frac{d}{dt}{V}(\textbf{k})&=&-\frac{i e^2Q_f^2}{4~m_f~k^0}\int d\textbf{q}
\:n_f(\textbf{x},\textbf{q})\Big[\Big(q\cdot\theta \cdot\ep_1~q\cdot\ep_1 - q\cdot\theta \cdot\ep_2~ q\cdot\ep_2\Big)~U(\textbf{k})~~~~~~\\ \nonumber
&-& \Big(q\cdot\theta \cdot\ep_2~q\cdot\ep_1 +q\cdot\theta \cdot\ep_1~ q\cdot\ep_2\Big)~Q(\textbf{k})\Big].
\label{rho01}
\end{eqnarray}
Using the matrix elements of the density operator, the time variation of $V$ in NC space can be cast into:
\begin{eqnarray}
&&
\dot{V}(\textbf{k})=i\frac{3}{4}\frac{\sigma^T}{\alpha~k^0}\,\frac{m_e^2}{\Lambda^2}\frac{\bar{\ep}_f}{g_f}(CQ+DU),
\label{vdot}
\end{eqnarray}
where
\begin{eqnarray}
&&
 C= -\epsilon^{ijk}~\hat{q}_i \hat{v}_k\Big(\ep_{1j}~\hat{q}\cdot\varepsilon_{2} + \ep_{2j}~\hat{q}\cdot\varepsilon_{1}\Big) ,
 \nonumber
\\
&&
 D= \epsilon^{ijk} \hat{q}_i~\hat{v}_k\Big( \ep_{1j}~\hat{q}\cdot\varepsilon_{1} - \ep_{2j}~\hat{q}\cdot\varepsilon_{2}\Big),
  \label{constant}
\end{eqnarray}
and $\hat{q}=\frac{\vec{q}}{|q|}$ is the direction of fermion beam, $m_e$ is the mass of electron, $\sigma^T$ is the Thomson cross section, $\alpha=e^2/4\pi$ and $\bar{\epsilon}_f$ is the energy
 density of fermion beam.\par
  In the laboratory frame, we set the laser-photon momentum $\mathbf{k}$ in the $\hat{z}$-direction (the direction of
incident laser beam), then the polarization vectors are defined as:
\begin{equation}\label{epsilon}
\varepsilon_1(k)=\hat{x},~\varepsilon_2(k)=\hat{y}.
\end{equation}
 We may also consider the direction of muon beam, $\hat{q}$ in the $x-z$ plane, $\hat{q}=sin\theta \hat{i}+cos\theta \hat{k}$, then for the space-like component of noncommutative parameter one has
\begin{eqnarray}
 \Delta \phi &=& \frac{\Delta V}{2 Q}
 \simeq  \frac{3}{8}\frac{\sigma^T}{\alpha~k^0}\,\frac{m_e^2}{\Lambda_{\tiny{NC}}^2}\frac{\bar{\ep}_{f}(\mathbf{x},\bar{\mathbf{q}})}{g_f}~~\Delta t~~ %
  (\frac{sin 2 \theta}{2}- {sin^2{\theta}})~\hat{v}.\hat{z}\nonumber\\
&\approx& 4.3\times 10^{-36}{\rm cm}^2\left(\frac{1~TeV}{\Lambda_{\tiny{NC}}}\right)^2 \left(\frac{\bar{\ep}_{f}(\mathbf{x},\bar{\mathbf{q}})}{k^0}\right)~ \Delta t~(\frac{sin 2 \theta}{2}- {sin^2{\theta}}),~~~~~
  \label{vss}
  \end{eqnarray}
  where $\Delta t$ is the time interval of the laser beam interacting with the charged lepton beam.
\section{Charged lepton beams}\label{lepton-beam}

In the previous section we found how the circular polarization of photons depend on the beam parameters in the NC-space.  Before calculating the Faraday conversion in fermion-photon scattering we briefly introduce the present available charged lepton beams. The charged particles suffering energy lost as the synchrotron radiation during the acceleration. However, for a particle with mass $m$ and beam energy $E$ in circular motion with radius $R$, the energy loss per revolution is given by $\Delta E\varpropto \frac{1}{R}(\frac{E}{m})^4$ which is smaller for a larger radius or for a more massive particle. Therefore, accelerating the electron as the lightest charged particle to very high energy is a difficult task in the circular collider.  In the LEP
experiment, electrons were accelerated up to $E=100~GeV$\cite{Lep}.  Meanwhile, in a linear collider such as an International Linear Collider(ILC) one can reach to higher energies about $E=1 TeV$ or even more.\par
 Furthermore, in laser-plasma accelerators, electron can be accelerated to energies from hundreds of MeV \cite{el1,el2,el3} to multi-GeV energies \cite{el4,el5}.\par
  Muon as the 2nd lightest charged particle in nature but not stable, is almost 200 times more massive than the electron. In fact, the muon mean lifetime is very short about $\sim 2.2~\mu s$, but enough to provide an intense beam which can be  accelerated to high energy.  However, pion decay can be usually used as a source for the muon production via $\pi^+ \rightarrow \mu^{+} +\nu_{\mu}$ and
  $\pi^- \rightarrow \mu^{-} +\bar{\nu}_{\mu}$.  Meanwhile, colliding high energy protons with nuclei produce  pions  through interactions such as:
 $p+p\rightarrow p+n+\pi^+$ and $p+n\rightarrow n+n+\pi^+$ for the proton energies more than 400 MeV and the double pion production in
  $p+p \rightarrow p+p+\pi^++\pi^-$  at the larger energies.  According to the momentum of the generated muon beam they are called "decay muon beam" or "surface muon beam".
The first type of muon beam is about 80 percent polarized and have momentum from 40 to several hundreds of MeV/c.
  Such muon beams are available at PSI, TRIUMF, J-PARC \footnote{Japan Proton Accelerator Research Complex} and RIKEN-RAL.
 The second one is often known as surface or Arizona beam\cite{arizona}
  that is 100 percent polarized and monochromatic and has kinetic energy 4.1 MeV.
  Such muon beams are available at PSI \footnote{Swiss Muon Source $S\mu S$}, TRIUMF, J-PARC, ISIS and RIKEN-RAL.\par
 Besides, very low energy muon beams (ultra slow muons with energy at eV-KeV range) can be generated by reducing the energy of an
  Arizona beam\cite{slowmuon} which is available in the PSI while the High-intensity low-energy muon beam is developing in J-PARC. Recently a multi-TeV Muon Collider, so called "Muon accelerator program" (MAP), has been planed. The purpose of this program is to develop the concepts and technologies required for the Muon Collider and Neutrino Factories \cite{MAP}.
  \section{Faraday conversion in Photon-Fermion beam interaction on NC space-time}
   For a typical muon beam with the number of muon per bunch $\sim 10^{12}$, the energy range  $\bar E_\mu\approx |\bar{\mathbf{q}}|\sim {\rm 1GeV-1TeV}$, the size of beam in the interaction region $\sim 1cm$ and the bunch length about $\sim 1cm$, one can estimate an average energy of flux per bunch as
\begin{equation}\label{Edensity}
\bar \varepsilon_\mu(\mathbf{x},\bar{\mathbf{q}})\approx |\bar{\mathbf{q}}|\,n_\mu(\mathbf{x},\bar{\mathbf{q}})c\sim 10^{12} \,{\rm TeV}/({\rm cm}^2{\rm s}).
\end{equation}
 This type of muon beam has an angle divergence about $\theta_{\rm div} \sim m_\mu/E_\mu$ which is negligible in the high energy muon beam accelerator which is the case for example in the MAP experiment\cite{MAP}.  To find the Faraday conversion for instance in Photon-Muon beams interaction we have for the interacting spot a spatial interval $\Delta d\sim 2 R_0+d.\theta_{\rm div}\sim {\rm 1cm}$ where d is the distance between muon beam and interaction spot with laser beam and a temporal interval of the order of $\Delta t\approx \Delta d/c\sim 10^{-10}\,$ sec.  Therefore, for a laser beam with energy of photons about  $k_0\sim 1eV$  the Eq.(\ref{vss}) for the conversion $\Delta \phi\Big|_{\mu B}$ leads to
 \begin{eqnarray}
 \Delta \phi\Big|_{\mu B}
 \approx 10^{-22}\left(\frac{1~TeV}{\Lambda_{\tiny{NC}}}\right)^2\left(\frac{\bar \varepsilon_\mu}{10^{12}{\rm TeV}{\rm cm}^{-2}{\rm s}^{-1}}\right)\left(\frac{\Delta t}{10^{-10}~{\rm s}}\right)\left(\frac{\rm eV}{k^0}\right)\sin^2{\theta},\label{muon0}
\end{eqnarray}
  where $\sin^2{\theta}\sim 1$ when the laser beam is perpendicular to the muon beam.  What is found in Eq.(\ref{muon0}) is the conversion for one interaction. However,  the conversion can be enhanced by multiple interaction of the laser beam with muons which can be usually provided by a setup of suitable mirrors.  If we suppose the
mirrors have reasonable absorption coefficient about $C_a\sim 10^{-5}$, the intensity of linearly polarized laser beam  $Q$  reduces to $Q \rightarrow Q(1- C_a)$ for each reflection. If we assume each laser pulse in its path can interact $N$-times with the muon beams, the maximum value of $N$ can be estimate as $\sum_{n=0}(1-C_a)^n\approx\frac{1}{C_a}$ \cite{Mohammadi}.  Therefore,  Eq.(\ref{muon0}) for the $N$-times reflections leads to
\begin{eqnarray}
\Delta \phi \Big|_{\mu B}
 \approx 10^{-17}\left(\frac{1~TeV}{\Lambda_{\tiny{NC}}}\right)^2\left(\frac{\bar \varepsilon_\mu}{10^{12}~{\rm TeV}{\rm cm}^{-2}{\rm s}^{-1}}\right)\left(\frac{\Delta t}{10^{-10}{\rm s}}\right)\left(\frac{\rm eV}{k^0}\right)\left(\frac{10^{-5}}{C_a}\right).\label{muon1}
\end{eqnarray}

 The Faraday conversion phase shift of a linearly polarized laser beam due to its interaction with the muon beam  for different values of  $\Lambda_{NC}$ is given in Tab.(\ref{tab:conversion1}).
\begin{table}[ht]
    \centering
   \caption{\small{$\Delta \phi \Big|_{\mu B}$ for linearly polarized laser beam due to its interaction with {\it muon beam} for different values of $\Lambda_{NC}$.}}
  \vspace{0.3cm}
   \scalebox{0.95}
  { \begin{tabular}{llll}
     \hline\hline
     $E_{\mu} (TeV)$&$\Lambda_{\tiny{NC}} (TeV)$&~~~$k_0(eV)$ ~~~&~~~~ $~~\Delta\phi_{\mu B}$\\
     \hline
    ~~~\raisebox{-1.5ex}{1}~~~&~~~\multirow{2}{*}{1}~~~&~~~1~~~& $~~~\sim 10^{-17}$\\
     ~~~&~~~&~~~0.1~~~& $~~~\sim 10^{-16}$\\
      \hline
      ~~~\raisebox{-1.5ex}{1}~~~&~~~\multirow{2}{*}{10}~~~&~~~1~~~& $~~~\sim 10^{-19}$\\
     ~~~&~~~&~~~0.1~~~& $~~~\sim 10^{-18}$\\
      \hline
      ~~~\raisebox{-1.5ex}{1}~~~&~~~\multirow{2}{*}{100}~~~&~~~1~~~& $~~~\sim 10^{-21}$\\
     ~~~&~~~&~~~0.1~~~& $~~~\sim 10^{-20}$\\
        \hline\hline \\
\end{tabular}} \label{tab:conversion1}
\end{table}
Meanwhile, one can use Eq.(\ref{muon1}) with appropriate changes for the other charged fermions.  For instance, for the electron one can consider the appropriate parameters from the LEP experiment\cite{Lep} in which the number of electrons per bunch is about $\sim 10^{11}$ locating at energies $\bar{E}_e\approx |\bar{\mathbf{q}}|\sim 100 {\rm GeV}$, where the size of beam in interaction region and bunch length are both about $\sim 1cm$. Therefore, the mean energy of flux of the electron beam can be estimated as $\bar\varepsilon_e\sim 10^{10}{\rm TeV}{\rm cm}^{-2}{\rm s}^{-1}$ which cast Eq.(\ref{muon1}) into
  \begin{eqnarray}\label{electron1}
\Delta \phi\Big|_{e B}
 \approx 10^{-19}\left(\frac{1~TeV}{\Lambda_{\tiny{NC}}}\right)^2\left(\frac{\bar \varepsilon_e}{10^{10}{\rm TeV}{\rm cm}^{-2}{\rm s}^{-1}}\right)\left(\frac{\Delta t}{10^{-10}{\rm s}}\right)\left(\frac{\rm eV}{k^0}\right)\left(\frac{10^{-5}}{C_a}\right).~~~
 \end{eqnarray}
In this case the values of $\Delta \phi\Big|_{e B}$ for different values of  $\Lambda_{NC}$ in interaction
 of linearly polarized laser beam and the electron beam can be obtained which is given in Tab.(\ref{tab:conversion2}).\par
 \begin{table}[ht]
   \centering
   \caption{\small {$\Delta \phi\Big|_{e B}$ for laser beam due to its interaction with {\it electron beam} for different values of $\Lambda_{NC}$.}
     }
  \vspace{0.3cm}
   \scalebox{0.95}{
   \begin{tabular}{llll}
     \hline\hline
     $E_{e} (GeV)$&$\Lambda_{\tiny{NC}} (TeV)$&~~~$k_0(eV)$~~~&~~~~$~~\Delta\phi_{eB}$\\
     \hline
    ~~~\raisebox{-1.5ex}{100}~~~&~~~\multirow{2}{*}{1}~~~&~~~1~~~&$~~~\sim 10^{-19}$\\
     ~~~&~~~&~~~0.1~~~& $~~~\sim 10^{-18}$\\
      \hline
      ~~~\raisebox{-1.5ex}{100}~~~&~~~\multirow{2}{*}{10}~~~&~~~1~~~&$~~~\sim 10^{-21}$\\
     ~~~&~~~&~~~0.1~~~& $~~~\sim 10^{-20}$\\
      \hline
      ~~~\raisebox{-1.5ex}{100}~~~&~~~\multirow{2}{*}{100}~~~&~~~1~~~&$~~~\sim 10^{-23}$\\
     ~~~&~~~&~~~0.1~~~& $~~~\sim 10^{-22}$\\
        \hline\hline \\
   \end{tabular}
   } \label{tab:conversion2}
 \end{table}
      \section{Generation Rate of circular polarization}
   In the previous section the Faraday conversion has been found to be at most $10^{-17}$ in the case of muons.  Although this value seems to be small we show that can be large enough for the rate of generating circular polarization for the laser beam interacting with the muon/electron beam. To this end, we use Eq.(\ref{vss}) in which $\Delta V (k)$ represents the number of generated circularly polarized photons with energy $|k| = k_0 \sim 1eV$ per unit area $(cm^2)$ per unit time $(s)$ for each laser pulse.
 The rate of the circular polarization generation for photons in the laser beam can be estimated as follows
     \begin{equation}\label{rate}
    R_{_V}\approx \left(\frac{\Delta V}{k^0}\right) \sigma_{\rm laser}\, f_{\rm pulse} \,  \, \tau_{\rm pulse},
\end{equation}
 where $\sigma_{\rm laser}$ is the laser-beam size which is smaller than the
charged lepton beam size $\Delta d$ and represents the effective area of photon-charged lepton interaction, $\tau_{\rm pulse}$ is the time duration of a laser pulse and $f_{\rm pulse}$ the laser repetition rate is the number of laser pulses per second. To have more efficiency, we assume that the laser and charged lepton beams are synchronized and the $f_{\rm pulse}$ is equal to $f_{\rm bunch}$ the repetition rate of beam which is the number of charged lepton bunches per second.  Therefore, one can easily find the value of $R_{_V}$ as follows
  \begin{eqnarray}
R_{_V} \simeq
10^{-16}f_{\rm pulse}N_\gamma\left(\frac{1~TeV}{\Lambda_{\tiny{NC}}}\right)^2\left(\frac{\bar \varepsilon_\mu}{10^{12}~{\rm TeV}{\rm cm}^{-2}{\rm s}^{-1}}\right)\left(\frac{\Delta t}{10^{-10}{\rm s}}\right)\left(\frac{\rm eV}{k^0}\right)\left(\frac{10^{-5}}{C_a}\right),\label{rate2}
\end{eqnarray}
where $N_\gamma$ is the number of photons per pulse which can be obtained as
\begin{equation}\label{rate3}
    N_\gamma= \frac{Q(k)}{k^0}\,\sigma_{\rm laser}\,\tau_{\rm pulse}= \frac{\varepsilon_{\rm pulse}}{k^0},
\end{equation}
and the total energy of a laser pulse $\varepsilon_{\rm pulse}=Q(k)\sigma_{\rm laser}\,\tau_{\rm pulse}$. Furthermore, the averaged power of a linearly polarized laser beam is approximately given by $ \bar{P}_{\rm laser}= f_{\rm pulse}\,\varepsilon_{\rm pulse}$ which cast Eq.~(\ref{rate2}) for muon into
\begin{eqnarray}
R_{_V}\Big|_{\mu B}\simeq
 10^{-16}\frac{\bar{P}_{\rm laser}}{k^0}\left(\frac{1~TeV}{\Lambda_{\tiny{NC}}}\right)^2\left(\frac{\bar \varepsilon_\mu}{10^{12}~{\rm TeV}{\rm cm}^{-2}{\rm s}^{-1}}\right)\left(\frac{\Delta t}{10^{-10}{\rm s}}\right)\left(\frac{\rm eV}{k^0}\right)\left(\frac{10^{-5}}{C_a}\right).\label{rate4}
\end{eqnarray}
In fact, for a muon beam with $\bar {\varepsilon}_{\mu}\sim 10^{12}\,{\rm TeV}\,{\rm cm}^{-2}\,{\rm s}^{-1}$, $\Lambda_{\tiny{NC}}=1TeV$ and a linearly polarized laser beam of energy $k^0\sim $ 1eV and the average power $\bar{P}_{\rm laser}\simeq 30~W$, the generation rate of the circularly polarized photons is about $R_{_V}\Big|_{\mu B} \sim 10^4/{\rm s}(\sim 3\times10^{11}/{\rm year}$). This rate seems to be large enough to be measured experimentally. The generation rates of the circular polarized photons $R_{_V}\Big|_{\mu B}$ for different values of noncommutative scale are given  in Tab.(\ref{tab:rate1}).\par
   \begin{table}[ht]
   \centering
   \caption{\small{The generation rate of circular polarization due to {\it muon} and laser beams interaction for different values of $\Lambda_{NC}$.}
     }
  \vspace{0.3cm}
   \scalebox{0.95}{
   \begin{tabular}{llll}
          \hline\hline
          \vspace{-0.35cm}
          \\
            $ E_{\mu}$ (TeV)&$\Lambda_{\tiny{NC}} (TeV)$&~~~~$R_V(1/s)$&~~~$\bar{P}_{\rm laser}({\rm KW})$\\
        \hline
     $~~~~1$~~~&~~~$1$ ~~~&~~~$\sim 10^{4}$&$~~~\sim 10^{-2}$\\
      $~~~~1$ ~~~&~~~$10$~~~&~~~$\sim 10^{4}$&$~~~\sim 1$\\
      $~~~~1$ ~~~&~~~$100$~~~&~~~$\sim 10^{4}$&~~~$\sim 10^{2}$\\
     \hline\hline \\
   \end{tabular}
   } \label{tab:rate1}
 \end{table}
 We can also give an estimation on the generation rate of the circular polarization due to the electron and laser beam interaction $R_{_V}\Big|_{eB}$ as follows
  \begin{eqnarray}
R_{_V}\Big|_{eB}\simeq
 10^{-18}\frac{\bar{P}_{\rm laser}}{k^0}\left(\frac{1~TeV}{\Lambda_{\tiny{NC}}}\right)^2\left(\frac{\bar \varepsilon_e}{10^{10}{\rm TeV}~{\rm cm}^{-2}{\rm s}^{-1}}\right)\left(\frac{\Delta t}{10^{-10}{\rm s}}\right)\left(\frac{\rm eV}{k^0}\right)\left(\frac{10^{-5}}{C_a}\right)\,.\label{rate5}
\end{eqnarray}
To obtain the rate $R_{_V}\Big|_{eB}\simeq 10^{4}/{\rm sec}$ for an electron beam with $\bar \varepsilon_e\simeq 10^{10}{\rm TeV}~{\rm cm}^{-2}{\rm s}^{-1}$ and $\Lambda_{\tiny{NC}}=1TeV$, a laser beam with average $\bar{P}\sim {\rm10^{-3}MW}$ is needed which seems to be available in the near future with the present laser technologies \cite{laser}.  Nevertheless,
 the generation rate of circular polarized photons due to the electron and laser beams interaction for different values of noncommutative scale and the laser average power is given in Tab.(\ref{tab:rate2}).\par
   \begin{table}[ht]
   \centering
   \caption{\small{The generation rates of the circular polarization due to {\it electron} and laser beams interaction for different values of $\Lambda_{NC}$.}
     }
     \vspace{0.3cm}
   \scalebox{0.95}{
   \begin{tabular}{llll}
     \hline\hline
      \vspace{-0.35cm}
          \\
     $E_{e} (GeV)$&$\Lambda_{\tiny{NC}} (TeV)$&~~~~$R_V(1/s)$&~~~$\bar{P}_{\rm laser}({\rm MW})$\\
     \hline
     $~~~~100$~~~&~~~$1$ ~~~&~~~$\sim 10^{4}$&$~~~\sim  10^{-3}$\\
      $~~~~100$ ~~~&~~~$10$~~~&~~~$\sim 10^{4}$&$~~~\sim 10^{-1}$\\
      $~~~~100$ ~~~&~~~$100$~~~&~~~$\sim 10^{4}$&~~~$\sim 10$\\
     \hline\hline \\
   \end{tabular}
    } \label{tab:rate2}
 \end{table}
\section{Conclusion}
We have considered the interaction of photons in an intense linearly polarized laser beam with ultra intense beams of muon and electron in the NC space time in the NCQED framework. Consequently, we have obtained the Faraday conversion phase shift $\Delta\phi$ and the generation rate $R_{_V}$ of the circularly polarized photons in the interaction with muon/electron beam which are given in  Eqs.(\ref{muon1}),(\ref{electron1}) and  Eqs.(\ref{rate4}),(\ref{rate5}), respectively. We have discussed the possibility of using advanced laser and muon/electron beams to probe the noncommutative effects in such processes even for the NC scale as large as $10 TeV$ with the available present technologies.  In fact,
 for a muon/electron beam with a flux intensity of $\bar\varepsilon_{\mu,e}\sim 10^{12}/10^{10}\,{\rm TeV}\,{\rm cm}^{-2}\,{\rm sec}^{-1}$ and a linearly polarized laser beam of energy $k^0\sim 1$eV and an average power $\bar{P}_{\rm laser}\simeq 1$ MW, the rate of generating of circularly polarized photons is about $R_{_V} \sim 10^4/{\rm sec}$ for a noncommutative energy scale about $\Lambda_{NC}\sim 10$TeV [see Tabs.(\ref{tab:conversion1}),(\ref{tab:conversion2}) for $\Delta\phi$ and Tabs.(\ref{tab:rate1}),(\ref{tab:rate2}) for $R_{_V}$]. This rate seems to be large enough to be measured experimentally in near future.

\end{document}